\documentclass[preprint,review,12pt]{elsarticle}

\usepackage{amssymb}
\usepackage{float}
\usepackage{color}
\usepackage{booktabs}
\usepackage{array}
\usepackage{tikz}
\usepackage{steinmetz}
\usepackage{multirow}
\usepackage{graphicx}
\usepackage{tabularx}
\usepackage{makecell} 
\usepackage{subcaption}
\usepackage{amsmath,amssymb,amsfonts}
\usepackage{algorithmic}
\usepackage{graphicx}
\usepackage{textcomp}
\usepackage{xcolor}
\usetikzlibrary{shapes.geometric, arrows.meta, positioning, fit}
\newcolumntype{P}[1]{>{\centering\arraybackslash}p{#1}}
\newcolumntype{M}[1]{>{\centering\arraybackslash}m{#1}}

\journal{Sustainable Energy, Grids and Networks}
\begin{document}

\begin{frontmatter}

\title{Data-Driven Reduction of Fault Location Errors in Onshore Wind Farm Collectors}

\author[eesc]{Alves Junior, A. J.\corref{cor1}}\ead{alailtonjunior@usp.br}
\author[eesc]{Davi, M. J. B. B.}
\author[eesc]{Fernandes, R. A. S.}
\author[eesc]{Oleskovicz, M.}
\author[eesc]{Coury, D. V.}
\address[eesc]{Department of Electrical and Computing Engineering, S\~ao Carlos School of Engineering, University of S\~ao Paulo, S\~ao Carlos, S\~ao Paulo, 13566-590, Brazil}
\cortext[cor1]{Corresponding author}

\begin{abstract}
Accurate fault location is essential for operational reliability and fast restoration in wind farm collector networks. However, the growing integration of inverter-based resources changes the current and voltage behavior during faults, challenging the effectiveness of traditional phasor-based diagnostic methods. In this context, the present paper introduces an advanced machine-learning solution that enhances a deterministic fault distance estimator by incorporating a correction model driven by a Gated Residual Network, specifically designed to minimize residual fault location errors. Through comprehensive feature engineering and selection processes, an improved predictor was developed and trained on a diverse set of fault scenarios simulated in a PSCAD-based real-world wind farm model, including variations in fault type, resistance, location, inception angle, and generation penetration. Hyperparameter optimization was performed using the Optuna framework, and the robustness of the method was statistically validated. Results show a significant improvement in accuracy, with a 76\% overall decrease in fault location error compared to state-of-the-art approaches. The proposed method demonstrates strong scalability and adaptability to topological and operational changes. This approach advances the deployment of data-driven fault location frameworks for modern power systems.
\end{abstract}


\begin{highlights}
\item Development of an integrated data-driven workflow that improves classic phasor-based fault location techniques with an ML-driven error correction component.
    
\item Comprehensive and in-depth analysis of feature engineering, encompassing both feature extraction and selection, to improve model performance and robustness.

\item Rigorous benchmarking using statistical validation, revealing both superior accuracy and stability of the proposed solution compared to the state-of-the-art methods.
\end{highlights}


\end{frontmatter}

\section{Introduction}

The accelerating global transition towards cleaner and more sustainable energy systems has led to wider integration of renewable sources, notably wind and solar, into modern power grids. This paradigm shift has placed Inverter-Based Resources (IBRs) at the forefront of the energy landscape, prompting new opportunities for decarbonization while introducing significant operational and technical challenges for power system engineers~\cite{IEEE2018}. The unique dynamic behaviors of IBR-dominated systems challenge the effectiveness of traditional fault diagnosis techniques, demanding efforts for adaptation and innovation~\cite{IEEE2018}.

Fault location methodologies, which are the focus of this paper, play an important role in ensuring reliable network operation and facilitating prompt service restoration. Among the diverse array of techniques reported in the literature, phasor-based methods stand out due to their direct applicability in existing monitoring infrastructures~\cite{DAS2014}. One-terminal schemes, encompassing impedance~\cite{ZIEGLER2011}, reactance~\cite{CAPAR2014}, and Takagi-based methods~\cite{TAKAGI1982} (including some modifications~\cite{DAS2014,SEL2018}), are widely referenced. Two-terminal strategies, in turn, rely on synchronized measurements from both ends of the protected line section, as reported by Girgis~\cite{GIRGIS1992}, Johns and Jamali~\cite{JJ1990}, Preston and Radojevic~\cite{PR2011}, and He~\cite{HE2011}. However, the growing penetration of IBRs has exposed limitations in these classical approaches, given the divergent fault current characteristics and control strategies inherent to modern converter-based generation~\cite{IEEE2018}, opening up an extensive field for new investigations and solution proposals.

Recent research addressing the intersection of IBR impacts and fault location has been concentrated on distribution networks~\cite{MAT2019,CHANG2022,Khattak} and IBR interconnection transmission lines~\cite{DAVI2023MULTI,DAVI2024110366}. For example, using energy storage to compensate IBR-induced fault current behaviors has been shown to improve fault identification in feeders~\cite{MAT2019}, while two-ended measurement strategies have been proposed specifically for microgrids with high inverter shares~\cite{CHANG2022}. Deep learning architectures, such as those outlined in~\cite{Khattak}, have demonstrated the ability to robustly classify and locate faults within increasingly complex, decentralized networks. In the context of transmission systems, particularly regarding IBR interconnection lines, both phasor- and Machine Learning (ML)-based proposals have emerged, offering dynamic solutions tailored to the prevailing IBR fault contributions~\cite{DAVI2023MULTI,DAVI2024110366}.

Although these advancements indicate progress, most methodologies remain rooted in the domains of IBR interconnection lines or distribution systems. As a result, fundamental differences in topology, configuration, and operational nuances inherent to onshore wind farm collector systems tend to be overlooked. Among the limited body of work that directly addresses this gap, the authors of~\cite{DAVI2025MULTI} offer an important contribution by assessing conventional phasor-based fault location algorithms within wind farm collectors and proposing a combined strategy to enhance overall accuracy. Nevertheless, particular challenges, such as increased errors for double-phase and three-phase faults, persist within this approach.

These gaps highlight the need for novel methodologies capable of delivering high-accuracy fault location across all fault types and under diverse operational scenarios, as the complexity of power grids intensifies with the proliferation of IBRs. In parallel, rapid advancements in ML have unlocked new potential for addressing complex and nonlinear problems in modern power systems, including fault location~\cite{DAVI2024110366}. In this context, the main contributions of this paper are summarized as follows:

\begin{itemize}
    \item Development of an integrated data-driven workflow that improves classic phasor-based fault location techniques with an ML-driven error correction component.
    
    \item Comprehensive and in-depth analysis of feature engineering, encompassing both feature extraction and selection, to improve model performance and robustness.

    \item Rigorous benchmarking using statistical validation, revealing both superior accuracy and stability of the proposed solution compared to the state-of-the-art methods.
\end{itemize}

The remainder of this paper is organized as follows: Section II details the proposed data-driven methodology, covering the overall approach, feature engineering, network architecture, and assessment strategy. Section~\ref{sec:test_system} introduces the wind farm collector test system and outlines the simulated fault scenarios. Section IV presents the results, including a comprehensive performance evaluation and a comparative analysis with existing techniques. Section V discusses the extended evaluation and practical considerations for implementation. Finally, Section VI concludes the paper by summarizing the key findings.

\section{Data-Driven Approach for Fault Location Error Reduction}

\subsection{Objective}

This paper proposes a data-driven methodology to improve the accuracy of fault location in wind farms. The main aim is to complement a conventional deterministic fault distance estimator with a ML model trained to predict a correction factor for the initial estimate. Following each fault event, the recorded data were used to refine the model, enabling a continuous learning process that improves predictive accuracy over time.

\begin{figure}[!ht]
	\centering\includegraphics[width=0.7\linewidth]{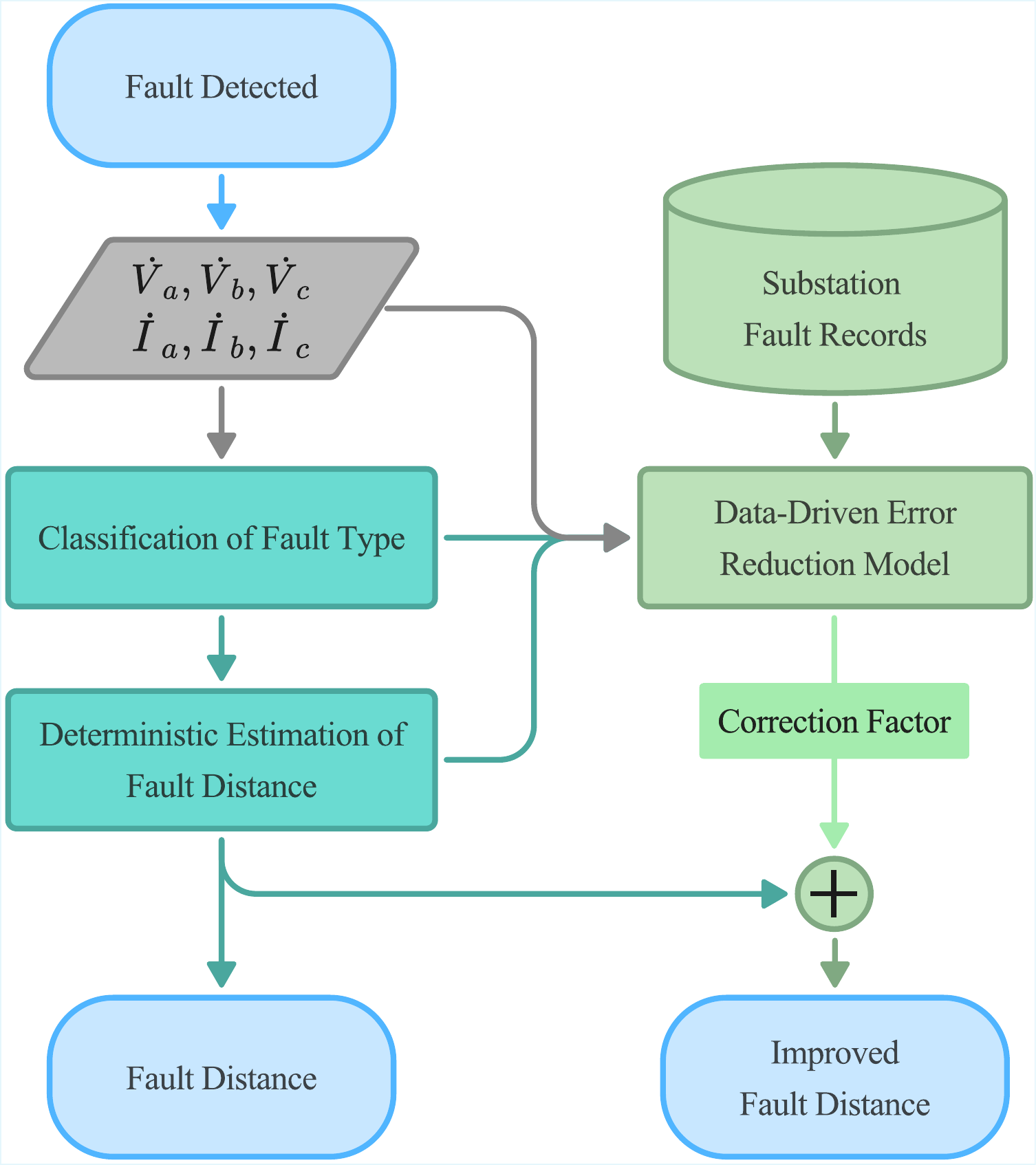}
	\caption{Proposed methodology for data-driven reduction of fault distance estimation errors.}\label{fig:methodology}
\end{figure}

The proposed workflow, illustrated in Fig.~\ref{fig:methodology}, begins with a baseline estimator providing an initial fault distance. This estimate, together with voltage and current phasor measurements from the event, is supplied to the Data-Driven Reduction Model. The model outputs a correction factor, which is added to the initial distance to produce a more accurate fault location. Through iterative refinement, the model progressively reduces location errors as additional fault data becomes available.

\subsection{Strategy for Feature Engineering}

Feature engineering plays a crucial role in the development of effective data-driven models, as appropriate feature design improves predictive performance and accelerates learning. 

\subsubsection{Feature Extraction}

The initial feature set for this study was constructed from variables directly related to or derived from fault distance estimation. Although the methodology is independent of the baseline fault locator, this study adopts the Multi-Method (MM) estimator proposed in \cite{DAVI2025MULTI}, whose formulation is expressed in Equation~\eqref{eq:multi-method}.

\begin{equation}
    d = \frac{Im[\dot{V}_{loop} \cdot \dot{I}_F^*]}{Im[Z_{L1} \cdot \dot{I}_{loop} \cdot \dot{I}_F^*]}\label{eq:multi-method}
\end{equation}

\noindent where $\dot{V}_{\text{loop}}$ and $\dot{I}_{\text{loop}}$ denote the loop voltage and current, respectively, $\dot{I}_F^*$ is the complex conjugate of the fault current, and $Z_{L1}$ is the positive-sequence line impedance. The loop quantities are defined according to the fault type, as shown in Table~\ref{tab:loopquantities}.

\begin{table}[!ht]
	\centering
    \setlength{\tabcolsep}{12pt}
	\caption{Multi-Method Quantities~\cite{DAVI2025MULTI}.}
	\label{tab:loopquantities}
	\begin{tabular}{cccc}
		\hline
		Fault Loop & $\dot{V}_{loop}$ & $\dot{I}_{loop}$ & $\dot{I}_{F}$ \\\hline
		$AG$ & $\dot{V}_{a}$ & $\dot{I}_{a}$ + $K_0$$\dot{I}_{0}$ & $\dot{I}_{0}$\\
		$AB$ & $\dot{V}_{a}$-$\dot{V}_{b}$ & $\dot{I}_{a}$-$\dot{I}_{b}$ & $\dot{I}_{2}\cdot e^{j30^\circ}$\\
        $ABG$ & $\dot{V}_{a}$+$\dot{V}_{b}$ & $\dot{I}_{a}$+$\dot{I}_{b} + 2K_0$$\dot{I}_{0}$ & $\dot{I}_{0}$\\
        $ABC$ & $\dot{V}_{a}$-$\dot{V}_{b}$ & $\dot{I}_{a}$-$\dot{I}_{b}$ & $\dot{I}_{a}$-$\dot{I}_{b}$\\
		\hline
	\end{tabular}
\end{table}

The initial candidate features were derived from voltage and current phasor measurements and grouped into four categories, as shown in Table~\ref{tab:features}.

\begin{table}[!ht]
\centering
\caption{Feature Groups Used for Fault Distance Estimation.}
\label{tab:features}
\resizebox{\textwidth}{!}{
\begin{tabular}{m{4.5cm}m{9cm}}
\toprule
\textbf{Group} & \textbf{Features} \\
\midrule
\textbf{Phasor Magnitudes and Angles} &
$|\dot{I}_a|$, $|\dot{I}_b|$, $|\dot{I}_c|$, $|\dot{V}_a|$, $|\dot{V}_b|$, $|\dot{V}_c|$, $\cos\left(\phase{\dot{I}_a}\right)$, $\cos\left(\phase{\dot{I}_b}\right)$, $\cos\left(\phase{\dot{I}_c}\right)$, $\cos\left(\phase{\dot{V}_a}\right)$, $\cos\left(\phase{\dot{V}_b}\right)$, $\cos\left(\phase{\dot{V}_c}\right)$, $\sin\left(\phase{\dot{I}_a}\right)$, $\sin\left(\phase{\dot{I}_b}\right)$, $\sin\left(\phase{\dot{I}_c}\right)$, $\sin\left(\phase{\dot{V}_a}\right)$, $\sin\left(\phase{\dot{V}_b}\right)$, $\sin\left(\phase{\dot{V}_c}\right)$\\
\midrule
\textbf{Symmetrical \phantom{312321} Components} &
$|\dot{I}_0|$, $|\dot{I}_1|$, $|\dot{I}_2|$, $|\dot{V}_0|$, $|\dot{V}_1|$, $|\dot{V}_2|$, 
$\Delta \dot{I}_0$, $\Delta \dot{I}_1$, $\Delta \dot{I}_2$, 
$\Delta \dot{V}_0$, $\Delta \dot{V}_1$, $\Delta \dot{V}_2$, 
$|\dot{I}_1^{\text{pre}}|$, $|\dot{V}_1^{\text{pre}}|$, $\cos\left(\phase{\dot{I}_0}\right)$, $\cos\left(\phase{\dot{I}_1}\right)$, $\cos\left(\phase{\dot{I}_2}\right)$, $\cos\left(\phase{\dot{V}_0}\right)$, $\cos\left(\phase{\dot{V}_1}\right)$, $\cos\left(\phase{\dot{V}_2}\right)$, $\sin\left(\phase{\dot{I}_0}\right)$, $\sin\left(\phase{\dot{I}_1}\right)$, $\sin\left(\phase{\dot{I}_2}\right)$, $\sin\left(\phase{\dot{V}_0}\right)$, $\sin\left(\phase{\dot{V}_1}\right)$, $\sin\left(\phase{\dot{V}_2}\right)$\\
\midrule
\textbf{Loop and Fault \phantom{3121} Features} &
$|\dot{I}_F|$, $|\dot{V}_{\text{loop}}|$, $|\dot{I}_{\text{loop}}|$, 
$|\dot{I}_{\text{loop}}^{\text{pre}}|$, 
$|\dot{V}_{\text{loop}}^{\text{pre}}|$, 
$\sin\left(\phase{\dot{V}_{\text{loop}}^{\text{pre}}}\right)$, $\cos\left(\phase{\dot{V}_{\text{loop}}^{\text{pre}}}\right)$, 
$\sin\left(\phase{\dot{I}_{\text{loop}}^{\text{pre}}}\right)$, $\cos\left(\phase{\dot{I}_{\text{loop}}^{\text{pre}}}\right)$, $\cos\left(\phase{\dot{I}_{\text{loop}}}\right)$,  $\cos\left(\phase{\dot{V}_{\text{loop}}}\right)$, $\sin\left(\phase{\dot{I}_{\text{loop}}}\right)$,  $\sin\left(\phase{\dot{V}_{\text{loop}}}\right)$,\\
\midrule
\textbf{Estimated fault \phantom{3123} distance and line \phantom{3123} impedance} &
$d$, $d_{\text{max}}$, $|Z_1|$, $\text{Re}(Z_1)$, $\text{Im}(Z_1)$  \\
\bottomrule
\end{tabular}}
\end{table}

In Table~\ref{tab:features}, subscripts $a$, $b$, and $c$ refer to the three-phase values; subscripts $0$, $1$, and $2$ denote the zero-, positive-, and negative-sequence symmetrical components, respectively. The operator $\Delta$ represents the variation of a phasor with respect to its pre-fault value, while the superscript “pre” indicates pre-fault conditions. The subscript $loop$ designates the loop quantities defined in Table~\ref{tab:loopquantities}. Finally, $d$ is the estimated fault distance, $d_{\text{max}}$ is the line length, and $Z_1$ is the positive-sequence line impedance, where $\text{Re}(Z_1)$ and $\text{Im}(Z_1)$ denote its real and imaginary parts.

\subsubsection{Feature Selection}

The construction of a suitable feature set required a systematic selection process to identify variables that provide high predictive value while avoiding redundancy. To this end, two complementary techniques were applied: Mutual Information (MI) \cite{Kraskov_2004} and Pearson correlation analysis.

First, MI was employed to quantify the statistical dependence between each candidate feature and the target variable, defined as the correction factor for fault location. Features with higher MI values exhibit stronger relevance to the prediction task. In parallel, Pearson correlation coefficients were computed to assess the degree of linear dependence between features. When two or more features exhibited correlation coefficients exceeding $|0.95|$, the set was considered highly collinear. In such cases, only the feature with the highest MI score was retained. 

\begin{figure}[!ht]
    \centering
    \includegraphics[width=0.75\linewidth]{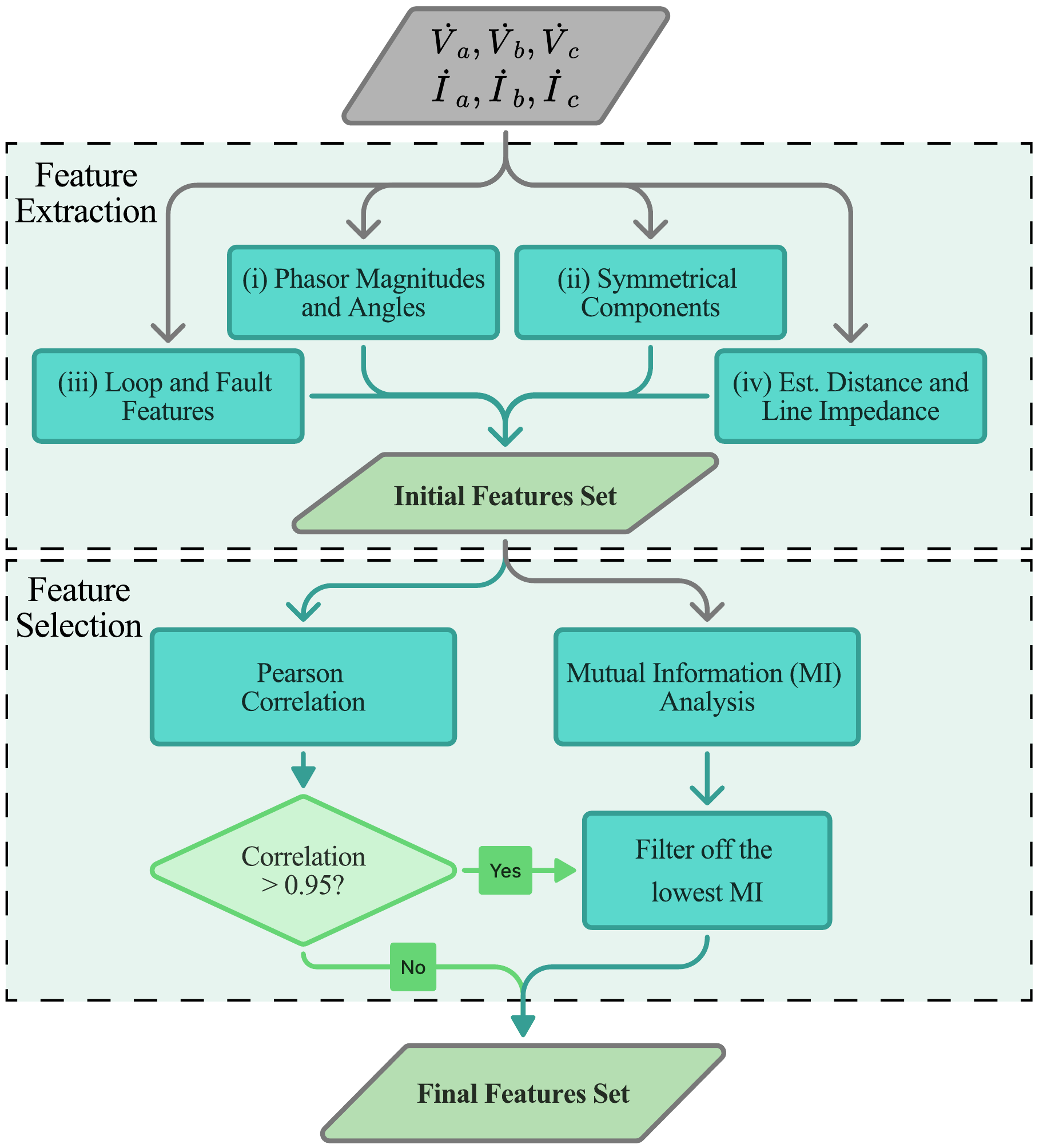}
    \caption{Overview of the Feature Engineering Process.}\label{fig:feature_selection}
\end{figure}

This combined procedure, shown in Fig.~\ref{fig:feature_selection}, ensures that the final feature set maximizes predictive information while minimizing redundancy. By retaining features with high MI and low interdependence, the model is expected to generalize more reliably to unseen fault scenarios while maintaining robustness in real-world applications.

\subsubsection{Feature Preprocessing Strategy}

Following feature selection, the final feature set underwent a preprocessing pipeline to ensure numerical stability and optimize model performance. Continuous numerical features were standardized using Z-score normalization, as defined in Equation~\eqref{eq:zscore}. This transformation rescales each feature to have zero mean and unit variance, preventing variables with large magnitudes from disproportionately influencing model training~\cite{hastie2009elements}. The mean ($\mu$) and standard deviation ($\sigma$) were computed exclusively from the training data to avoid information leakage from the validation set.

\begin{equation}
x' = \frac{x - \mu}{\sigma}
\label{eq:zscore}
\end{equation}

The single categorical feature, \texttt{Fault Type}, was processed using one-hot encoding. Since this variable is nominal with no inherent order, the encoding converts it into a set of binary values, one per fault category. This representation allows the model to interpret fault type without imposing artificial ordinal relationships, ensuring independent treatment of all categories.

\subsection{Gated Residual Network Architecture for Error Reduction}

\begin{figure}[!ht]
\centering
\includegraphics[width=0.8\linewidth]{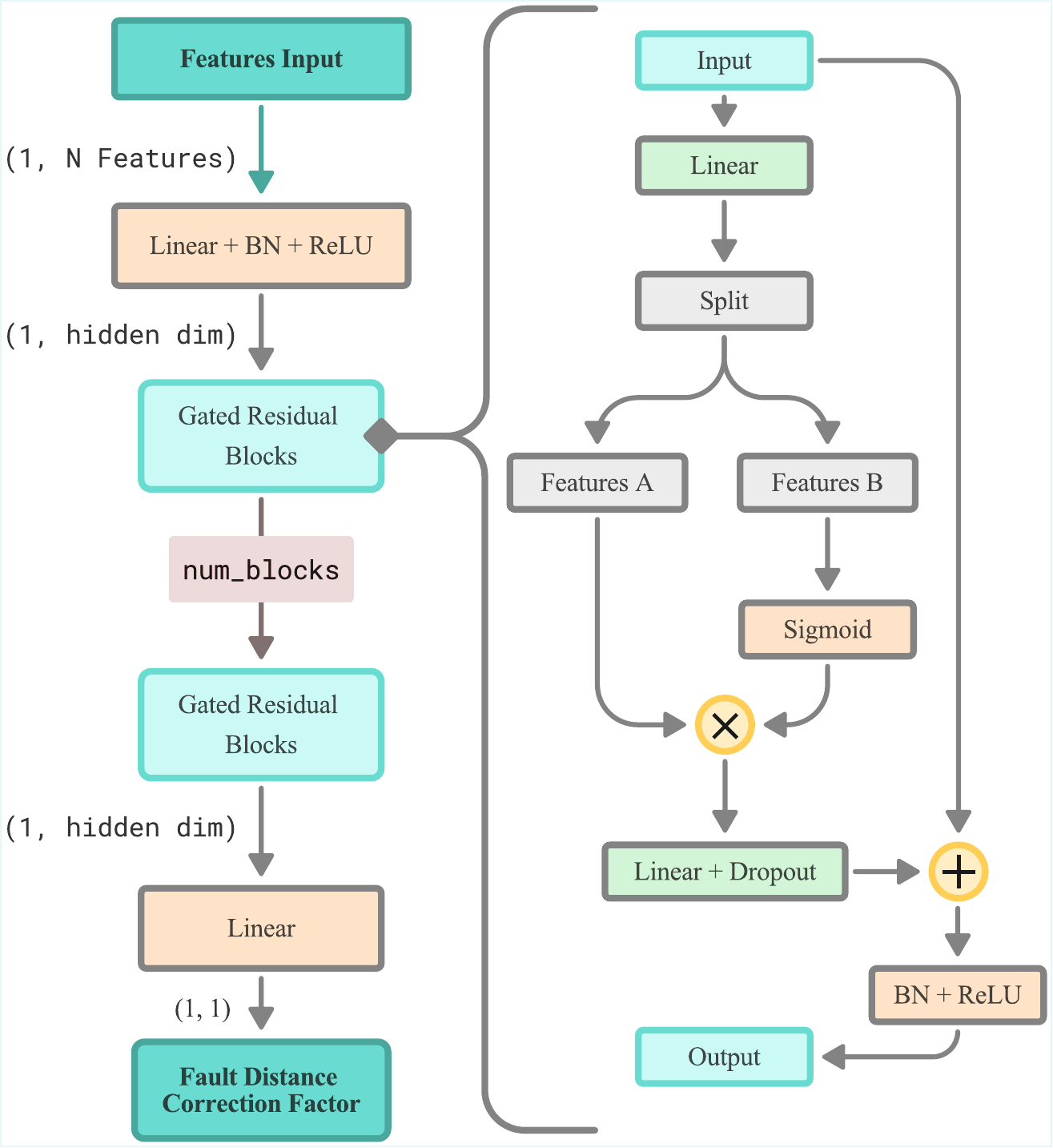}
\caption{Architecture of the GRN model for Error Reduction.}\label{fig:ann_model}
\end{figure}

The proposed correction model is a deep neural network based on the Gated Residual Network (GRN) architecture~\cite{savarese2016learningidentitymappingsresidual}, defined in this work as GRN-MM, as the Multi-Method fault locator was employed as the baseline. The overall architecture is illustrated in Fig.~\ref{fig:ann_model}.

The model consists of three main components: an input projection layer; a stack of Gated Residual Blocks (GRBs); and a final regression output layer. Each stage progressively transforms the input features into increasingly abstract representations while maintaining numerical stability and mitigating overfitting.

The input projection layer maps the raw input feature vector into a hidden representation of dimension \texttt{hidden\_dim} via a fully connected linear layer, followed by batch normalization and a ReLU (Rectified Linear Unit) activation. This combination enhances representational capacity, standardizes feature distributions, and introduces non-linearity, preparing the features for deeper processing.

The core of the architecture was formed by a stack of GRBs. Each GRB integrates a Gated Linear Unit~\cite{dauphin2017languagemodelinggatedconvolutional}, which adaptively regulates information flow using a learned gating mechanism (Fig.~\ref{fig:ann_model}). This gating mechanism allows the network to selectively emphasize or suppress feature components, improving the model's ability to capture complex dependencies between variables. To enhance stability during training, each GRB also incorporates residual connections, which facilitate gradient propagation across layers and mitigate the vanishing gradient problem. In addition, batch normalization and dropout were applied to further improve generalization and reduce overfitting.

The refined hidden representation produced by successive GRBs was mapped to a single scalar output by the regression layer, representing the corrective adjustment applied to the initial fault location estimate.

\subsubsection{Model Training Parameters}

The training process employed the Mean Absolute Error (MAE) as the loss function, defined in Equation~\eqref{eq:mae_loss}. MAE was selected for its robustness to outliers and its direct interpretability in regression tasks~\cite{Qi_2020}.

\begin{equation}
\text{MAE} = \frac{1}{n} \sum_{i=1}^{n} |y_i - f(x_i)|,
\label{eq:mae_loss}
\end{equation}
where $y_i$ is the ground-truth value, $f(x_i)$ is the model prediction, and $n$ denotes the total number of samples. Compared to squared-error-based losses, MAE reduces the influence of large deviations, making it more robust to outliers in the training data.

Model optimization was carried out using the AdamW algorithm~\cite{loshchilov2019decoupledweightdecayregularization}, which extends Adam by decoupling weight decay from the gradient-based update rule, thereby enhancing convergence stability. The parameter update at iteration $t$ is expressed in Equation~\eqref{eq:adamw}.

\begin{equation}
\theta_{t+1} = \theta_t - \eta \left( \frac{\hat{m}_t}{\sqrt{\hat{v}_t} + \epsilon} + \lambda \theta_t \right),
\label{eq:adamw}
\end{equation}

\noindent where $\theta$ represents model parameters, $\eta$ the learning rate, $\hat{m}_t$ and $\hat{v}_t$ the bias-corrected first and second moment estimates, $\lambda$ the weight decay coefficient, and $\epsilon$ a small constant for numerical stability.

\subsection{Methodology for Assessing the GRN-MM Performance}

The GRN-MM was optimized and assessed through a twofold procedure encompassing hyperparameter optimization and statistical validation. Hyperparameter optimization refined the model configuration for improved accuracy, while statistical validation verified the consistency and robustness of the results across different simulation conditions.

\subsubsection{Hyperparameter Optimization Approach}

Optimal model configuration was determined via automated hyperparameter optimization using the Optuna framework~\cite{TPE}. Optuna employs the Tree-structured Parzen Estimator (TPE), a Bayesian optimization strategy that efficiently explores high-dimensional parameter spaces by prioritizing regions likely to yield improved performance. This probabilistic approach is advantageous when training deep networks, where parameter interactions are often complex and non-linear.

The GRN-MM optimization consisted of 2,000 trials. Each trial evaluated a unique hyperparameter set via cross-validation on the training dataset, minimizing MAE. The search space covered network capacity, regularization, and optimization dynamics, summarized in Table~\ref{tab:GRN-MM_hyperparams}.

\begin{table}[!ht]
\centering
\caption{Search space parameters used in GRN-MM tuning via Optuna.}
\label{tab:GRN-MM_hyperparams}
\begin{tabular}{lllc}
\toprule
\textbf{Parameter} & \textbf{Description} & \textbf{Search Space} \\
\midrule
\texttt{hidden\_dim} & Hidden layer width & \{64, 128, ..., 1,024\} \\
\texttt{num\_blocks} & Number of residual blocks & [2, 10]\\
\texttt{dropout} & Dropout rate & [0.1, 0.6] \\
\texttt{lr} & Learning rate & [$10^{-4}$, $10^{-2}$]\\
\bottomrule
\end{tabular}
\end{table}

\subsubsection{Statistics Evaluation Approach}

The statistical stability of the optimized GRN-MM model was assessed by repeating the training and evaluation process 1,000 times. Each repetition was performed with a distinct random seed to account for the stochastic nature of network weight initialization and the training algorithm. This approach produces a distribution of the MAE on the test dataset, rather than a single deterministic value.

This repeated evaluation is critical given the non-convex optimization landscape of neural networks. Different initializations may cause the training process to converge to distinct local minima, resulting in performance variations. By analyzing the outcomes of multiple independent runs, it can be determined whether the observed performance is a consistent property of the model's architecture or an outlier resulting from a favorable initialization.

The resulting metric distributions allow quantitative assessment of robustness. Narrow distributions with low variance indicate consistent convergence to high-quality solutions, reflecting stability. Wider distributions indicate sensitivity to initialization, suggesting potential fragility in deployment. Therefore, this statistical evaluation strengthens the validity of the experimental findings by ensuring that reported results are representative and not influenced by isolated, atypical training runs.

\section{Test System}
\label{sec:test_system}

\begin{figure}[!ht]
	\centering
    \vspace{-0.3cm}
	\includegraphics[width=0.8\linewidth]{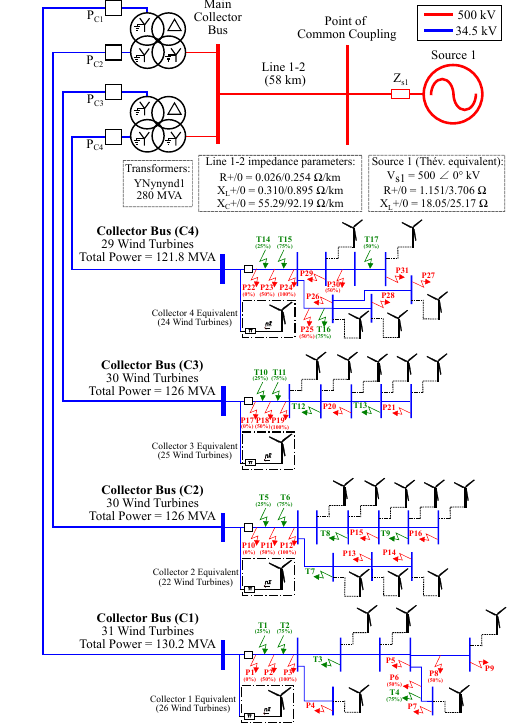}
	\caption{Test system parameters and single-line diagram.} 
	\label{fig:Sistema}
\end{figure}

A representative model of a real-world wind farm (60 Hz) utilizing IBRs was developed in the PSCAD software for this study. The single-line diagram, along with system details, can be seen in Fig.~\ref{fig:Sistema}. 

This arrangement comprises four principal collector busbars (C1, C2, C3, and C4) linked to 120 wind turbines equipped with full-converter technology. To conduct fault analysis within the collector network, one feeder per collector busbar was individually modeled in detail, while the other circuits were represented by equivalents~\cite{EQUIV2008}. Faults were simulated at multiple points to ensure comprehensive coverage: scenarios were placed along the main feeders at 0\%, 25\%, 50\%, 75\%, and 100\% of their lengths, as well as at various locations within each collector branch on cable segments ranging from 50 to 300 meters.

Several fault simulations were carried out at the pre-defined points indicated in Fig.~\ref{fig:Sistema}. To ensure a rigorous and unbiased evaluation, these locations were divided into two distinct sets: points labeled P1 through P31 were used to generate the training datasets, while points T1 through T17 were reserved exclusively for the testing dataset. These scenarios encompassed multiple fault types, including single-phase-to-ground (AG, BG, and CG), phase-to-phase (AB, BC, and CA), double-phase-to-ground (ABG, BCG, and CAG), and three-phase (ABC) faults, as well as a variation of fault resistance and fault inception angles. Different operating conditions were also considered, including distinct wind farm generation penetration levels, with the wind turbines operating at a unity power factor. The fault and generation parameters used to form the training/validation and test sets are specified in Table \ref{TabConjuntos}.
\vspace{-8pt}
\begin{table}[H]
	\centering
	\caption{Definition of simulated scenarios for training/validation and testing ML-based approach.}
	\label{TabConjuntos}
	\begin{tabular}{M{4cm}M{4cm}M{4cm}}		
		\hline	
		\textbf{Parameter} & \textbf{Training/Validation Stage Scenarios} & \textbf{Testing Stage Scenarios}
		\tabularnewline
		\hline
		Fault type & AG -  AB - ABG  - ABC &  AG - BG - CG - AB - BC - CA - ABG - BCG - CAG - ABC 
		\tabularnewline
		\hline
		Fault resistance & 0 $\Omega$ - 10 $\Omega$ - 25 $\Omega$ - 40~$\Omega$ - 50 $\Omega$ & 5 $\Omega$ - 15 $\Omega$ - 30 $\Omega$
		\tabularnewline
		\hline
		Fault inception angle & 0°  & 45° 
		\tabularnewline
		\hline
		Fault location Point & P1 to P31 & T1 to T17
		\tabularnewline		
		\hline
        Generation Level & 0.1 pu - 0.25 pu - 0.5~pu - 0.75 pu & 0.2 pu - 0.4 pu - 0.6~pu
		\tabularnewline		
		\hline
	\end{tabular}
\end{table}
\vspace{-26pt}
\section{Results for Assessing GRN-MM in Fault Location}
\vspace{-8pt}
\subsection{Results of the Feature Engineering}

Feature selection began by computing the MI score between each candidate feature and the target. The MI scores, sorted in descending order, are provided in Table~\ref{tab:mi_scores_sorted}. Higher MI values indicate stronger dependency on the target, suggesting greater predictive relevance.
\begin{table}[!ht]
\centering
\caption{MI scores for the input features with respect to the fault distance correction factor, sorted in descending order.}
\label{tab:mi_scores_sorted}
\resizebox{\textwidth}{!}{
\begin{tabular}{lc|lc|lc|lc}
\toprule
\textbf{Feature} & \textbf{MI} & 
\textbf{Feature} & \textbf{MI} & 
\textbf{Feature} & \textbf{MI} & 
\textbf{Feature} & \textbf{MI} \\
\midrule
$d$ & 1.8220 & $|I_F|$ & 0.9036 & $|I_2|$ & 0.7188 & $\Delta I_2$ & 0.7188 \\
$\Delta V_2$ & 0.7046 & $|V_2|$ & 0.7052 & $|V_\text{loop}|$ & 0.6172 & $|I_a|$ & 0.5203 \\
$|I_b|$ & 0.5700 & $|I_\text{loop}|$ & 0.5152 & $|I_0|$ & 0.5559 & $\Delta I_0$ & 0.5555 \\
$\Delta V_a$ & 0.5544 & $|I_c|$ & 0.5460 & $|V_a|$ & 0.5619 & $\Delta V_0$ & 0.5266 \\
$|V_0|$ & 0.5267 & $\Delta V_c$ & 0.4457 & $|V_c|$ & 0.4419 & $\Delta I_b$ & 0.4288 \\
$\Delta V_b$ & 0.4151 & $\Delta V_1$ & 0.4206 & $|V_1|$ & 0.4009 & $\Delta I_1$ & 0.4038 \\
$|V_b|$ & 0.3753 & $\Delta I_a$ & 0.3724 & $\cos(I_\text{loop})$ & 0.3703 & $\cos(I_b)$ & 0.3655 \\
$\cos(I_F)$ & 0.3586 & $\cos(I_a)$ & 0.3250 & $\cos(V_\text{loop})$ & 0.3245 & $\sin(I_1)$ & 0.3032 \\
$\cos(I_c)$ & 0.3016 & $\cos(V_2)$ & 0.2861 & $\sin(I_\text{loop})$ & 0.2822 & $\cos(I_1)$ & 0.2714 \\
$\cos(V_c)$ & 0.2628 & $\sin(I_F)$ & 0.2641 & $\sin(V_c)$ & 0.2554 & $\sin(I_c)$ & 0.2550 \\
$\sin(V_2)$ & 0.2300 & $\sin(V_\text{loop})$ & 0.2268 & $d_{\text{max}}$ & 0.2193 & $\cos(V_1)$ & 0.2155 \\
$\cos(V_b)$ & 0.1742 & $\cos(V_a)$ & 0.1725 & $\sin(I_2)$ & 0.1697 & $\sin(I_0)$ & 0.1690 \\
$\sin(V_\text{loop}^{\text{pre}})$ & 0.1656 & $\cos(V_\text{loop}^{\text{pre}})$ & 0.1623 & $\cos(I_\text{loop}^{\text{pre}})$ & 0.1783 & $\sin(I_\text{loop}^{\text{pre}})$ & 0.1577 \\
$\cos(I_0)$ & 0.1553 & $\sin(V_0)$ & 0.1169 & $|Z_1|$ & 0.1152 & $\text{Im}(Z_1)$ & 0.1155 \\
$\cos(V_0)$ & 0.1075 & $|I_\text{loop}^{\text{pre}}|$ & 0.1291 & $|V_\text{loop}^{\text{pre}}|$ & 0.0808 & $|I_1^\text{pre}|$ & 0.0637 \\
$|V_1^\text{pre}|$ & 0.0492 & $\sin(V_1)$ & 0.0567 & $\text{Re}(Z_1)$ & 0.0000 & & \\
\bottomrule
\end{tabular}}
\end{table}

To eliminate redundancy, features with strong linear correlations (Pearson coefficient $> |0.95|$) were grouped. Within each group, the feature with the highest MI was retained (Table~\ref{tab:feature_selection}). Features with MI scores below 0.1 were discarded, resulting in a final feature set that is highly informative, non-redundant, and suitable for training the GRN-MM model.

\begin{table}[!ht]
\centering
\caption{Selected features from highly correlated groups.}
\label{tab:feature_selection}
\begin{tabular}{l c | l c}
\hline
\makecell[l]{\textbf{Features}\\\textbf{in Group}} & \makecell{\textbf{Chosen}\\\textbf{Feature}} &
\makecell[l]{\textbf{Features}\\\textbf{in Group}} & \makecell{\textbf{Chosen}\\\textbf{Feature}} \\
\hline
$\cos(I_F)$, $\cos(I_\text{loop})$ & $\cos(I_\text{loop})$ & 
$\cos(I_\text{loop}^\text{pre})$, $\cos(V_\text{loop}^\text{pre})$ & $\cos(I_\text{loop}^\text{pre})$ \\
$\sin(I_F)$, $\sin(I_\text{loop})$ & $\sin(I_\text{loop})$ & 
$\sin(I_\text{loop}^\text{pre})$, $\sin(V_\text{loop}^\text{pre})$ & $\sin(V_\text{loop}^\text{pre})$ \\
Im($Z_1$), $d_{\text{max}}$, $|Z_1|$ & $d_{\text{max}}$ & 
$\Delta I_0$, $|I_0|$ & $|I_0|$ \\
$\Delta I_1$, $|I_1|$ & $|I_1|$ & 
$\Delta I_2$, $\Delta V_2$, $|I_2|$, $|V_2|$ & $|I_2|$ \\
$\Delta I_a$, $|I_a|$ & $|I_a|$ & 
$\Delta I_b$, $|I_b|$ & $|I_b|$ \\
$\Delta I_c$, $|I_c|$ & $|I_c|$ & 
$\Delta V_0$, $|V_0|$ & $|V_0|$ \\
$\Delta V_1$, $|V_1|$ & $\Delta V_1$ & 
$\Delta V_2$, $|I_2|$, $|V_2|$ & $|I_2|$ \\
$\Delta V_a$, $|V_a|$ & $|V_a|$ & 
$\Delta V_b$, $|V_b|$ & $\Delta V_b$ \\
$\Delta V_c$, $|V_c|$ & $\Delta V_c$ & 
$|I_2|$, $|V_2|$ & $|I_2|$ \\
\hline
\end{tabular}
\end{table}

\subsection{Results of the HyperTuning}

The Optuna-based optimization procedure identified a set of hyperparameters that effectively maximizes the performance of the GRN-MM model. The final configuration, summarized in Table~\ref{tab:hyperparams_results}, achieves an appropriate balance between representational capacity and regularization, thereby mitigating overfitting while maintaining high predictive accuracy.

\begin{table}[!ht]
\centering
\caption{Results of the hyperparameters tuning via Optuna.}
\label{tab:hyperparams_results}
\begin{tabular}{lcc}
\toprule
\textbf{Parameter} & \textbf{Best Parameter} & \textbf{Parameter Importance}\\
\midrule
\texttt{hidden\_dim} & 192 & 12\% \\
\texttt{num\_blocks} & 2 & 37\% \\
\texttt{dropout} & 0.1298 & 7\% \\
\texttt{lr} & 0.00828 & 44\% \\
\bottomrule
\end{tabular}
\end{table}

\begin{figure}[!ht]
\centering
\includegraphics[width=0.9\linewidth]{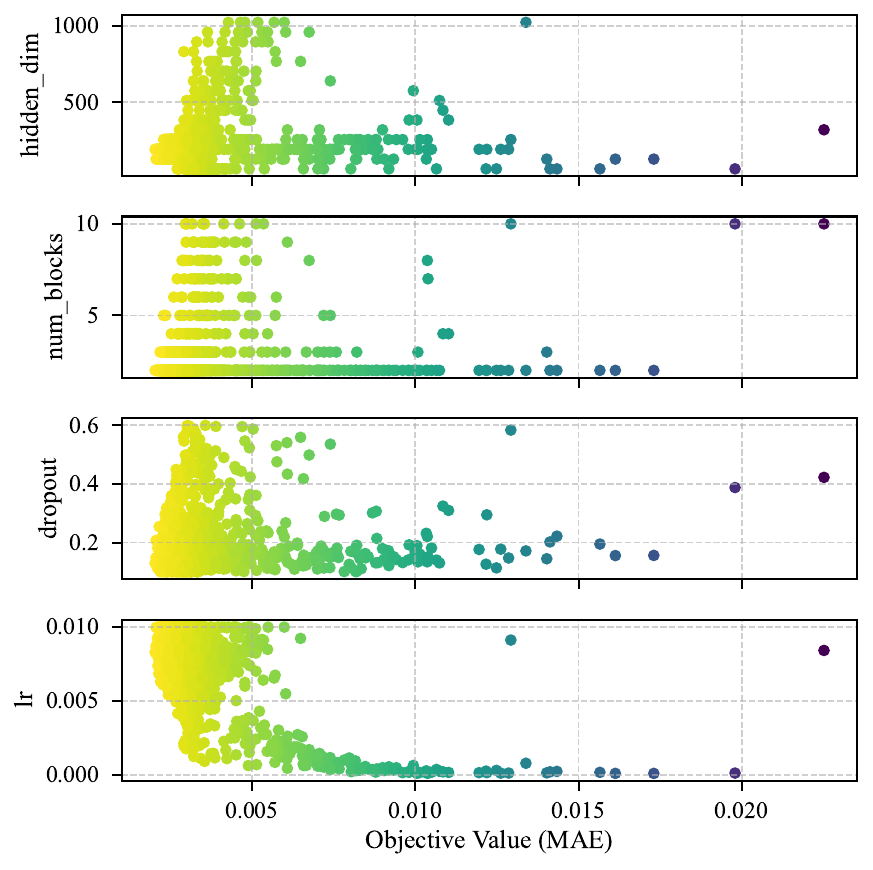}\vspace{-12pt}
\caption{Convergence of the objective function over 2,000 Optuna trials.}~\label{fig:placeholder}
\end{figure}

As indicated by the parameter importance scores in Table~\ref{tab:hyperparams_results}, the learning rate (\texttt{lr}) and the number of residual blocks (\texttt{num\_blocks}) were the most influential hyperparameters, accounting for 44\% and 37\% of the performance variability, respectively. The slice plot in Figure~\ref{fig:placeholder}, which visualizes the relationship between hyperparameters and the objective function (based on MAE) over 2,000 trials, illustrates this sensitivity. The best results were consistently obtained with learning rates within a narrow range (approximately 0.005 to 0.01). Similarly, models with 2 to 6 residual blocks generally outperformed deeper architectures, suggesting that excessive model depth may lead to overfitting.

In contrast, the hidden layer width (\texttt{hidden\_dim}) and dropout rate (\texttt{dropout}) exhibited less influence on model performance, as observed by their lower importance scores and the broader distribution of high-performing values in Figure~\ref{fig:placeholder}. This indicates that the model is less sensitive to the precise tuning of these parameters. The final optimized model employed a hidden dimension of 192, providing sufficient representational capacity for the task.

Overall, the hyperparameter optimization converged on a configuration featuring a low learning rate and a moderately sized architecture, a combination that yielded robust and accurate predictive performance.

\subsection{Statistic Evaluation of the Tuned Model}

\begin{figure}[!ht]
	\centering
    \vspace{-0.2cm}
	\includegraphics[width=1\linewidth]{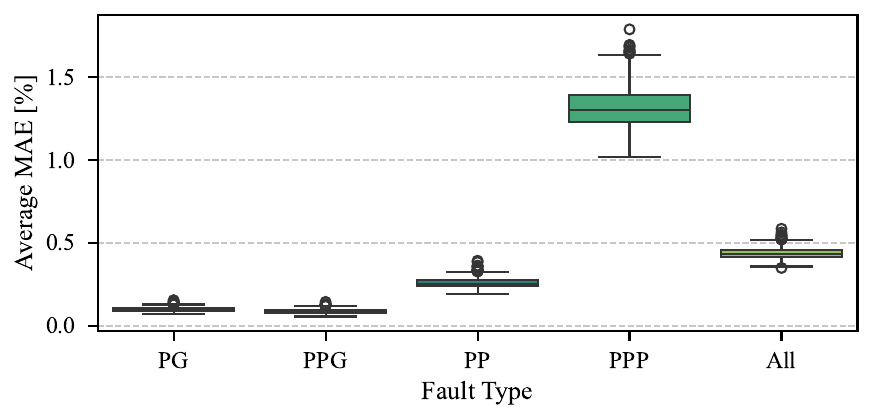}\vspace{-12pt}
	\caption{Distribution of the model's MAE by fault type across 2,000 distinct runs.}\vspace{-8pt}
	\label{fig:mae_distribution}
\end{figure}

The stability of the optimized GRN-MM model was evaluated by conducting 1,000 independent training and testing runs, each with a different random seed. Figure~\ref{fig:mae_distribution} presents the resulting distribution of MAE values for each fault type and for all faults combined (All). The box plots illustrate the median, interquartile range, and overall spread of the errors, offering a comprehensive view of the model's performance consistency.

The narrow MAE distributions obtained for phase-to-ground (PG), phase-to-phase-to-ground (PPG), and phase-to-phase (PP) faults highlight the stability and robustness of the proposed model across these categories. Although the three-phase (PPP) case displayed a slightly broader distribution, suggesting a marginally greater sensitivity to initialization, its median MAE remained low, reinforcing the overall reliability of the approach. Taken together, these results indicate that the model consistently converges to effective solutions regardless of random weight initialization, thereby minimizing the influence of stochastic factors in the training process and ensuring dependable performance across different fault scenarios.

\subsection{Proposed Fault Locator and Comparative Performance Analysis with Existing Methods}

To assess the effectiveness of the proposed model, a comparative analysis was conducted against both the baseline deterministic method and a set of well-established impedance-based fault location techniques. For consistency, the results from the median-performing run out of 1,000 independent training runs were selected. This choice avoids bias from extreme outliers and ensures that the reported performance reflects a representative outcome of the general GRN-MM trained model.

Fault location accuracy was quantified in terms of the percentage error relative to the line length, as defined in Equation~\eqref{eq:distance_error}:
\vspace{-5pt}
\begin{equation}
\text{Error} = \left| \frac{d_{\text{est}} - d_{\text{true}}}{d_{\text{max}}} \right| \times 100\%, \label{eq:distance_error}
\end{equation}

\noindent where $d_{\text{est}}$ is the estimated fault distance, $d_{\text{true}}$ is the actual fault distance, and $d_{\text{max}}$ is the total line length.

\begin{figure}[!ht]
	\centering
\includegraphics[width=1\linewidth]{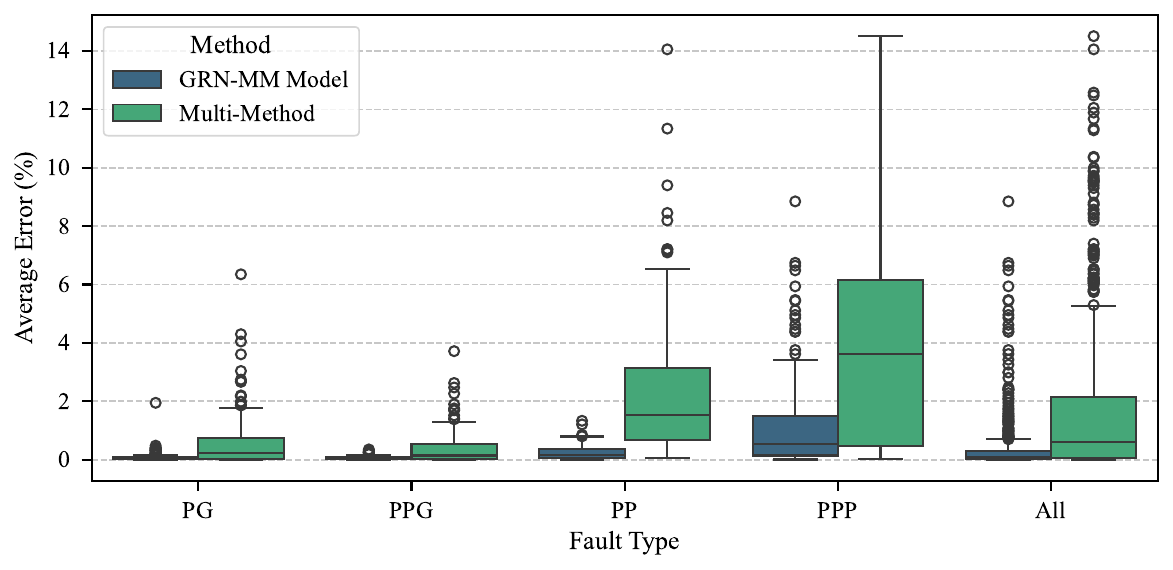}
	\vspace{-32pt}
	\caption{
	Distribution of absolute errors for the GRN-MM and the deterministic baseline}
	\label{fig:error_boxplot}
\end{figure}

Figure~\ref{fig:error_boxplot} compares the distribution of the distance errors for the proposed GRN-MM and the Multi-Method baseline.

The results of the comparative analysis indicate that the proposed GRN-MM model consistently outperforms the baseline Multi-Method. An evaluation of all fault types, as aggregated in the "All" category, shows the GRN-MM model achieves a lower median average error and exhibits considerably less variance compared to the baseline. This trend of improved performance is observed across the specific PG, PPG, and PP fault categories. The most significant difference between the two methods is evident in the analysis of three-phase (PPP) faults. For this fault type, the baseline method presents a high median error with a wide data distribution, whereas the GRN-MM Model maintains a low error rate and a much tighter distribution.

\begin{figure}[!ht]
	\centering
\includegraphics[width=1\linewidth]{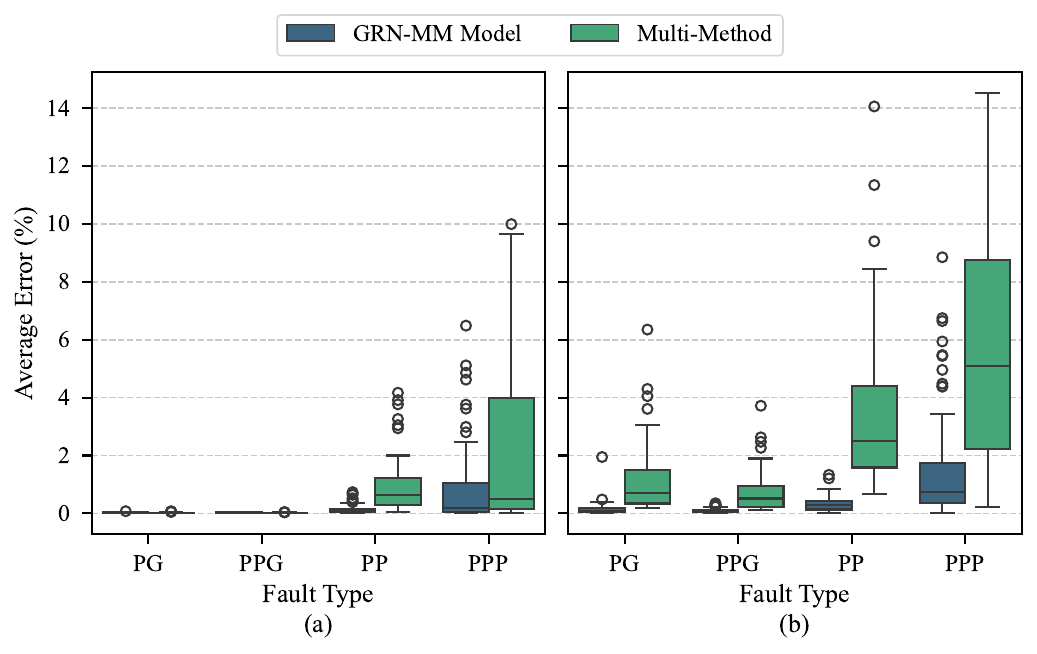}
	\vspace{-32pt}
	\caption{Distribution of average errors for the GRN-MM and the deterministic baseline, split by line type. (a) Main Overhead line and (b) other lines.}
\label{fig:error_boxplot_split_by_line_type}
\end{figure}

The performance of the baseline method was observed to be line-dependent, where fault location errors on the main overhead line were considerably lower than those on other collector lines. This discrepancy is illustrated in Figure~\ref{fig:error_boxplot_split_by_line_type}.

Although the baseline method achieved very low errors for PG and PPG faults on the main overhead line, the GRN-MM still outperforms it. This result indicates that the proposed model does not artificially raise the lower bound error values to improve median performance. Instead, it enhances the initial distance estimation while assigning a minimal or even zero correction factor to cases where the baseline already provides highly accurate fault location distances.

\begin{figure}[!ht]
\centering\includegraphics[width=1\linewidth]{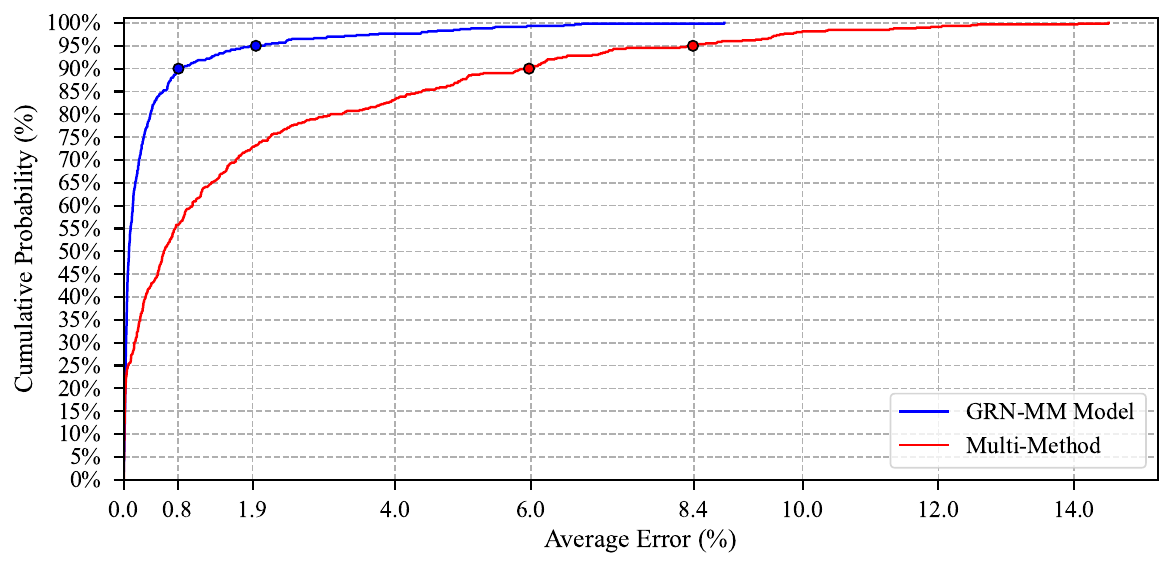}\vspace{-12pt}
	\caption{CDF of average errors for the median-performing run.}
	\label{fig:cdf_plot}
\end{figure}

The Cumulative Distribution Function (CDF) of the absolute errors, presented in Figure~\ref{fig:cdf_plot}, further highlights the performance improvement. The steeper slope of the GRN-MM curve indicates a higher concentration of predictions with low errors. For instance, the CDF shows that over 95\% of the GRN-MM's predictions have an error below 1.9\%, whereas the baseline method achieves this accuracy for only approximately 75\% of the fault cases.

Finally, Table~\ref{tab:final_results} summarizes the fault distance percentage errors for the proposed GRN-MM alongside several conventional methods and the Multi-Method baseline, divided by fault type.

\begin{table}[!ht]
\centering
\caption{Average percentage errors for all the fault scenarios.}
\label{tab:final_results}
\begin{tabular}{@{}lccccc@{}}
\toprule
\textbf{Method} & \textbf{PG} & \textbf{PPG} & \textbf{PP} & \textbf{PPP} & \textbf{ALL} \\
\midrule
IMPE \cite{ZIEGLER2011}   &     101\% &    408\% &  180\% &   104\% &    199\% \\
REAC \cite{CAPAR2014}  &      47.33\% &     25.25\% &   13.40\% &     3.96\% &     22.47\% \\
TAKS \cite{TAKAGI1982}  &       3.54\% &     24.62\% &    6.40\% &    13.43\% &     12.01\% \\
TAKN \cite{SEL2018} &      1.63\% &   1137\% &    2.30\% &  - &    -\\
TAKZ \cite{TAKAGI1982} &       0.59\% &  - &  - &  -  & - \\
$\text{TAKZ}_{\text{new}}$ \cite{DAVI2023MULTI}  &         -&      0.38\% &         -  &        -  &    - \\
Multi-Method \cite{DAVI2025MULTI} & 0.59\%  & 0.38\%  & 2.30\%  & 3.96\%  & 1.81\% \\\midrule
\textbf{GRN-MM } & \textbf{0.09\%}  & \textbf{0.07\%}  & \textbf{0.24\%}  & \textbf{1.34\%}  & \textbf{0.44\%}\\
\bottomrule
\end{tabular}
\end{table}

The GRN-MM consistently delivers the lowest error values in every category, achieving an overall error of only 0.44\%, which represents a significant reduction compared to the 1.81\% obtained with the Multi-Method baseline. Moreover, its improvements are evident not only in the aggregate performance but also across individual fault types, including the more challenging PPP cases. By contrast, traditional methods such as IMPE and REAC exhibit large errors, often exceeding 20\% or even several hundred percent in certain scenarios, underscoring their limitations in modern wind farm environments.

This consistent superiority demonstrates the effectiveness of the proposed hybrid framework. The proposed methodology not only enhances the baseline estimator but also adapts robustly to diverse fault conditions, confirming its potential as a reliable and scalable solution for practical fault location in wind farm collector systems.

\section{Extended Evaluation and Practical Considerations}

The proposed error correction framework was designed to be method-agnostic, enabling its integration with a variety of deterministic fault location algorithms. To illustrate this versatility, the methodology was first applied to a distinct estimator, thereby demonstrating its adaptability beyond the Multi-Method baseline. In addition to numerical improvements, it is also important to evaluate the practical considerations and limitations that accompany real-world deployment in IBR-rich wind farm collector systems. Therefore, this section presents (i) an application of the framework to an alternative fault locator, and (ii) a discussion on operational aspects that may affect field implementation.

\subsection{Application of the Methodology using another FL Method}

\begin{figure}[!ht]
\centering\includegraphics[width=1\linewidth]{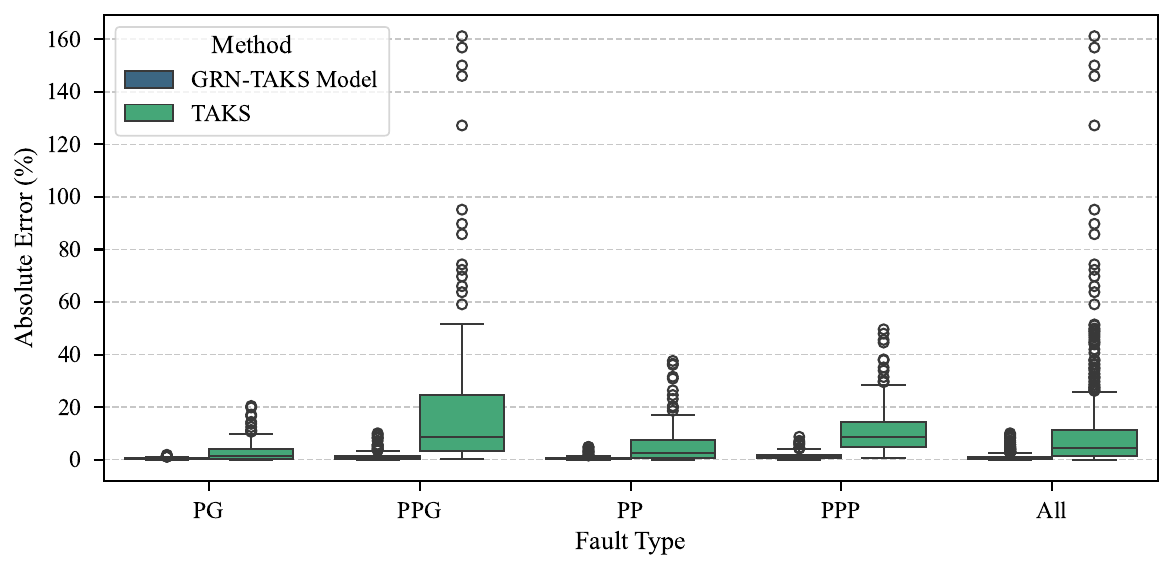}\vspace{-12pt}
	\caption{Distribution of average errors for the GRN-TAKS model and the standalone TAKS baseline method.}
	\label{fig:takz_plot}
\end{figure}

To further validate the versatility and method-agnostic nature of the proposed error correction framework, its performance was also assessed in combination with a different deterministic fault locator. In this case, the TAKS method was selected as the baseline estimator, and the resulting hybrid model is referred to as GRN-TAKS. The comparative results are illustrated in Figure~\ref{fig:takz_plot}.

As shown in Figure~\ref{fig:takz_plot}, the proposed correction framework substantially improves the performance of the TAKS method. The overall average error is reduced from 12.01\% to 1.0\%, corresponding to an improvement of 91.7\%. In addition, the disruptive maximum error observed in PPG faults, which reached 161\% with the standalone TAKS method, was reduced to only 10\% when applying the GRN-based correction. This represents a 93.8\% reduction in the extreme outlier values, highlighting the effectiveness of the proposed approach.

Importantly, the GRN-TAKS model not only outperforms the original TAKS method but also achieves superior results compared to the Multi-Method baseline, which was specifically developed as an enhanced estimator for wind farm collector systems. These findings reinforce the adaptability and effectiveness of the proposed methodology, demonstrating its capacity to enhance diverse deterministic fault location schemes and thereby offering a acceptable and scalable framework for fault distance estimation in modern power systems.

\subsection{Practical Considerations and Limitations of the Proposed Fault Locator}

Practical applications of advanced fault location methods in wind farm collector systems requires not only technical innovation, but also attention to operational conditions and system constraints. The data-driven approach developed in this work offers notable benefits in accuracy and adaptability, targeting challenges in IBR-rich grids. However, effective field integration calls for balanced consideration of both its strengths and the inherent characteristics of ML-based solutions.

Key practical aspects supporting real-world implementation of the proposed fault locator include the following:

\begin{itemize}
    \item Marked reduction in fault location errors across diverse scenarios, enhancing operational reliability and enabling faster restoration, essential in renewable power plants.
    \item Compatibility with existing measurement infrastructure, utilizing phasor data already present in modern substations.
    \item Flexible feature engineering and selection, allowing adaptation to different network configurations and scalability for a range of project sizes.
    \item Robustness promoted by comprehensive feature screening and statistical validation, ensuring stable performance even with generation level or fault parameter variations.
\end{itemize}

Nevertheless, it is also important to acknowledge certain inherent limitations typically associated with data-driven methodologies:

\begin{itemize}
    \item Initial access to a representative dataset, whether simulated or measured, remains as a prerequisite. Fortunately, this step is becoming increasingly feasible due to the growing sophistication of grid simulation tools and the widespread use of digital monitoring in wind farms. When real operational cases from the system are available, they provide highly valuable representative datasets. However, in such situations, building a sufficiently comprehensive dataset is possible but may take considerable time, as the data must be gradually accumulated under real operating conditions.
    \item As most data-driven methods, the proposed solution offers less interpretability than conventional methods, potentially needing supplementary reporting to foster operator trust and regulatory acceptance, particularly in highly regulated environments.    
\end{itemize}

In summary, the proposed fault locator demonstrates strong practical viability, combining accuracy, adaptability, and compatibility with the deployment realities of renewable energy systems. When supported by a robust data infrastructure, the methodology provides measurable benefits to wind farm operations, with manageable and well-characterized limitations that can be addressed through continued development and integration.

\section{Conclusions}

This paper presented a novel data-driven methodology for fault location in onshore wind farm collector systems. By integrating a conventional method with GRN-based architecture, significant improvements in fault location accuracy were achieved under a wide variety of fault conditions, collector circuit topologies, and generation level scenarios.

Comprehensive assessments using a real-world wind farm model demonstrated that the proposed approach reduced the average distance error by 76\% compared to state-of-the-art techniques. Robustness and scalability were confirmed through extensive feature engineering, rigorous statistical validation, and testing, showing stable performance across different fault types and the ability to adapt to varying network and generation profiles. The methodology also benefits from compatibility with existing measurement infrastructure.

Overall, the proposed solution advances the state-of-the-art in fault location applied to onshore wind farm collectors, offering both measurable performance benefits and compatibility with modern wind farm infrastructures. Future directions include field deployment, facilitating the transition towards smarter and more resilient renewable power grids.

\section*{Acknowledgments}

The authors would like to thank the Sao Paulo Research Foundation (FAPESP) [$\#$ 2024/17884-3] and the  Coordenação de Aperfeiçoamento de Pessoal de Nível Superior – Brasil (CAPES) for their financial support. We gratefully acknowledge the support of the RCGI – Research Centre for Greenhouse Gas Innovation, hosted by the University of São Paulo (USP), sponsored by FAPESP [$\#$2020/15230-5], and sponsored by TotalEnergies and the strategic importance of the support given by ANP (Brazil’s National Oil, Natural Gas and Biofuels Agency) through the R\&DI levy regulation.

\bibliography{ref.bib}

\end{document}